\PassOptionsToPackage{numbers}{natbib}
\documentclass[11pt]{article}

\usepackage[preprint]{neurips_2026}

\usepackage{times}
\usepackage{latexsym}
\usepackage[table]{xcolor}
\usepackage[T1]{fontenc}

\usepackage[utf8]{inputenc}
\newcommand{\compactcode}[1]{%
  \begin{minipage}[t]{0.9\columnwidth}
  \setstretch{0.75} %
  \lstset{basicstyle=\ttfamily\footnotesize} %
  \lstinline|#1|
  \end{minipage}
}

\usepackage{microtype}

\usepackage{inconsolata}

\usepackage{graphicx}

\usepackage{booktabs}

\usepackage{setspace}
\usepackage{multirow}

\usepackage{hyperref}

\usepackage{todonotes}
\usepackage{tabularx}
\usepackage{fancyvrb}
\usepackage{listings}

\newcommand{\smallinline}[1]{\lstinline[basicstyle={\ttfamily\tiny}]{#1}}
\usepackage{cprotect}

\definecolor{jsonkey}{RGB}{148,0,0}
\definecolor{jsonvalue}{RGB}{0,128,0}
\lstdefinelanguage{json}{
    basicstyle=\scriptsize\ttfamily\color{jsonvalue},
    breaklines=true,
    tabsize=2,
    showstringspaces=false,
    literate=
        *{"byte\_buffers"}{{{\color{jsonkey}"byte\_buffers"}}}{14}
         {"full\_nl\_response"}{{{\color{jsonkey}"full\_nl\_response"}}}{18}
         {"qa\_pairs"}{{{\color{jsonkey}"qa\_pairs"}}}{10}
         {"category"}{{{\color{jsonkey}"category"}}}{10}
         {"question"}{{{\color{jsonkey}"question"}}}{10}
         {"response"}{{{\color{jsonkey}"response"}}}{10},
}

\title{Large Byte Model: Teaching Language Models About Compiled Code}

\author{%
  \begin{minipage}[t]{0.30\textwidth}
    \centering
    \textbf{Florian Störtz} \\
    \textbf{Sandra Servia-Rodríguez} \\
    \normalfont CrowdStrike U.K.
  \end{minipage}%
  \hfill
  \begin{minipage}[t]{0.38\textwidth}
    \centering
    \textbf{Catalin-Andrei Stan} \\
    \textbf{Alexandru Dinu} \\
    \textbf{Mihaela Gaman} \\
    \textbf{Calin Miron} \\
    \normalfont CrowdStrike Romania
  \end{minipage}%
  \hfill
  \begin{minipage}[t]{0.25\textwidth}
    \centering
    \textbf{Edward Raff} \\
    \normalfont CrowdStrike USA
  \end{minipage}%
    \\ \\
\small{
    \textbf{Correspondence:} \href{mailto:florian.stoertz@crowdstrike.com}{florian.stoertz@crowdstrike.com}
    \vspace{-2em}
}}

\begin{document}
\maketitle

\begin{abstract}
Malware analysis starts with the raw bytes of an executable program, and tools to ``lift'' these to higher-level representations, such as assembly, are expensive and subject to error. Large Language Models (LLMs) cannot process raw byte representations and answer questions about them. To this end, we present the first byte-native LLM%
  . Based on a vocabulary expansion technique using a bespoke byte tokenizer, %
  such a model is capable of responding to complex questions about malware binaries, with accuracies ranging from 69\% for malware family classification to 98\% for architecture classification.
  Our findings indicate that providing domain knowledge during training is essential for this application -- off-the-shelf models lack both accuracy and insight. We've deployed this emerging solution to a limited number of analysts to gather feedback for further improvements.%
\end{abstract}

\section{Introduction}
\label{sec:Intro}

Since the advent of widely accessible generative AI, threat actors have harnessed Large Language Models (LLM) to generate sophisticated malware at unprecedented speed, including data destruction scripts and shell command injection. Additionally, LLM-aided vulnerability exploit development has been used by several threat groups, including Iran-nexus adversaries \cite{csthreat25}, and enabled by the recently published multi-agent framework Co-RedTeam \cite{he2026}, thanks to its integrated security domain knowledge and code-aware analysis.
Developing exploits essentially entails translating from a known-to-work format (text and code) to an opaque, unreadable format (compiled binary executable code), given current LLM capabilities. 

Malware analysis involves understanding the behavior and purpose of a malware sample to prevent future cyberattacks, and starts with the opaque bytes of executable code. Malware attribution, explainability, and metadata extraction are some of the necessary steps for this analysis. An internal survey of our company's threat analysts revealed that most of these steps involve cumbersome manual processes. Manual review includes answering questions such as \textit{How many sections does this file have?}, \textit{What type of imports is it using?}, and \textit{Are there hints of process injection?}. The survey also revealed that metadata aggregation with at-a-glance summaries (labels, malware families, confidence scores) and highlighting critical sections of the file for the analyst to focus on are among the analysts' key needs. The surge in malicious code driven by LLM-based tools, often presented in unique variants that evade traditional signature-based detection, requires threat analysts to do this manual work at scale.

In order to assist threat analysts with malware analysis, we introduce Large Byte Model (LBM)\footnote{We first presented results from this model at the Fal.Con conference in September 2024, see https://www.crowdstrike.com/en-us/blog/ai-innovation-spotlight-falcon-2024/}, a byte-native Large Language Model that has gained knowledge about the intricacies of binary files and is still able to answer questions in natural language. %
Our model architecture represents a significant advancement in combining natural language processing with binary analysis. LBM achieves this through a hybrid embedder system that combines text and byte embedders, thereby extending an existing LLM's vocabulary with specialized byte tokens. The training process occurs in two %
phases: a pre-training phase that establishes fundamental byte-level capabilities, followed by a refinement phase that maintains language reasoning while developing binary analysis skills. In the first phase, we learn a new vocabulary of +5k byte tokens by training a BPE algorithm on 10GB of raw binary data. In the second, we perform instruction-based fine-tuning, using text and byte elements for both instruction and response, and incorporating expert knowledge about the byte buffer into the text element. We use several optimizations for long-context training, including Fully Sharded Data Parallel for multi-node multi-GPU training and Deep Speed’s sequence parallelism algorithm, to handle binary sequences of up to 256KB per chunk.
We have developed benchmarks to quantify the model's semantic conception of binary files at the levels of byte-opcode embeddings and model perplexity, providing lower-level insight than instruction-based tasks.

This model is currently able to perform some of the key tasks identified in the survey, including binary architecture assessment, malicious/benign binary classification, behavior analysis (such as exfiltration detection), and the identification of significant sections within binaries, with accuracies ranging from 69\% malware family classification to 98\% for architecture classification. We have deployed our trained model internally using AWS SageMaker.
Although further improvement is needed, we establish a scientific result that LLMs can be post-hoc improved to address a complex multimodal task that differs significantly from classical vision or audio tasks. It is highly relevant in the face of threat actors' increasing use of AI capabilities to generate and obfuscate malware.

\section{Related Work}
\label{sec:Related}

Previous works have applied LLM architectures, %
trained primarily on text, %
to byte information \cite{yu2023,gemini24,wu2024,pagnoni2024}. Contrary to our work, these are byproducts of general-purpose LLM models. They have not ingested vast amounts of dedicated byte data%
, nor do they incorporate expert threat analyst knowledge through fine-tuning.
One such example is Google's Gemini \cite{gemini24}. Our model differs from Gemini in two major aspects. Firstly, Gemini is not able to process raw binary data, and relies on decompilation and disassembly, which are costly steps to perform at scale. Instead, we process raw large byte files along with natural language queries and generate both natural language and byte data in a cost-effective way. Secondly, while Google’s approach leverages Gemini’s very large context size, we propose an efficient byte tokenizer that processes large binary files and compresses byte streams to fit a given model's context window. Furthermore, our solution is at least an order of magnitude cheaper than the estimated cost of developing Gemini \cite{aiind24}.

Other works, like byteGPT \cite{wu2024}, BLT (Byte Latent Transformer) \cite{pagnoni2024}, and MEGABYTE \cite{yu2023}, treat all input data modalities (including text, image, audio etc.) at the byte level and focus on byte-level next token prediction. However, %
their performance on text generation is very low when compared to %
existing LLMs. Instead, our novel approach builds on the existing LLM's ability to generate human-like language and integrates new knowledge about byte data via a vocabulary-expansion technique, thereby treating byte data as a separate text modality.%
This is possible thanks to an efficient tokenization method for byte data, powered by a combination of the large amounts of (un)labeled binary data that we have at our company, and our in-house malware analysis expertise. Although the concept of vocabulary expansion of LLMs is not novel \cite{kim2024}, to our knowledge, it has only been applied to combine languages with disjoint character sets, not to bridge the gap between differing modalities (in our case, bytes and text).

\section{Dataset Composition}
\label{sec:Dataset}
\subsection{Real-world Data}

We assembled a training and testing dataset from PE, MachO, and ELF binaries of 16 and 32 bit architectures, containing %
malware sampled across malware families and clean files. We leave 64-bit architectures for a future iteration, as these often involve larger files. Packed files, i.e., files that show signs of compression or encryption, were excluded from the dataset due to increased entropy (noise), rendering a byte-level analysis close to impossible. %
To remain within the base model's context length window, we split each file's byte data into 128 KB chunks (or "byte buffers"). On average, one file contains 3.5 byte buffers.
Our current sampling yields %
631,759 data points (see Table \ref{tab:dataset_comp}).

\begin{table}
\centering
\begin{tabular}{|l|r|r|r|}
\hline
\textbf{Label Distribution} & \textbf{Train} & \textbf{Test} & \textbf{Total} \\
\hline
Malicious & 216,280 & 54,126 & 270,406 \\
Benign & 165,952 & 41,256 & 207,208 \\
Adware & 123,175 & 30,970 & 154,145 \\
\hline
\textbf{Total Number of Files} & \textbf{505,407} & \textbf{126,352} & \textbf{631,759} \\
\hline
\hline
Number of buffer chunks & 1,786,358 & 448,342 & 2,234,700 \\
Total byte buffer size & 45.15 GB & 11.34 GB & 56.49 GB \\
\hline
\end{tabular}
\vspace{1em}
\caption{Real-world dataset composition.}
\label{tab:dataset_comp}
\end{table}

We enriched this collection %
with metadata, e.g., filetype, malware vs clean vs adware attribution (label), malware families, threat types, signatures extracted from dynamic analysis (sandbox). First, each sample was parsed using the radare2 reverse engineering framework~\cite{radare2} to extract generic file information (file type, architecture, bits, compiler information, etc.)
and bytes from the \textit{executable} sections, which we refer to as binary sequences%
. Secondly, metadata elements were assembled into coherent natural-language sentences using predefined templates. A training instance sample is shown in \autoref{lst:datasample}.

\begin{figure}[!h]
    \centering
\begin{lstlisting}[
    language=json,
    basicstyle=\scriptsize\ttfamily,
    breaklines=true,
    tabsize=2,
    showstringspaces=false,
    mathescape=false,
    literate={_}{\_}1
]
{"byte_buffers": [
    b"U\x8b\xec\x83\xe4..."
],
"full_nl_response": "A pe, 32 bits, x86 arch (i386 machine) file has been analyzed and is considered malicious. It is linked to the tilde malware family, known for engaging in ransomware activities.  The file displays the following characteristics: Creates a writable file in a temporary directory; Queries kernel debugger information; ...",
"qa_pairs": [
{
  "category": "fileinfo",
  "question": "Could you summarize the file characteristics of this byte sequence?",
  "response": "The provided file has been determined to be of pe, 32 bits, x86 arch (i386 machine) format."
},
{
  "category": "label",
  "question": "Can you assess the threat level of this byte buffer?",
  "response": "I have analyzed the file and determined it to be malicious. It appears to be part of the tilde malware family, which is associated with a ransomware threat."
},
{
  "category": "sandbox_signatures",
  "question": "What are the behavioral characteristics of this byte buffer?",
  "response": "The sandbox environment revealed the following behaviors: Creates a writable file in a temporary directory; Queries kernel debugger information; ..."
}]}
\end{lstlisting}
    \caption{Example of a training instance in our framework. Note that \texttt{byte\_buffers} and \texttt{sandbox\_signatures} are truncated here due to length.}
    \label{lst:datasample}
\end{figure}

The dataset is a combination of 300k files of file size $\leq$ 32kb with only architecture and benign/malicious/adware information (low metadata richness), and 331k files of file size $\leq$ 512kb with richer metadata information, such as malware family and sandbox signatures (see Table \ref{tab:dataset_metadata}). 

\begin{table}
\centering
\begin{tabular}{|l|r|r|r|}
\hline
\textbf{Metadata Fields} & \textbf{Train} & \textbf{Test} & \textbf{Total} \\
\hline
fileinfo,label & 240,203 & 59,797 & 300,000 \\
+malware\_family & 7,753 & 1,961 & 9,714 \\
+threat\_type & 238,773 & 59,824 & 298,597 \\
+sandbox\_signatures & 2,304 & 617 & 2,921 \\
fileinfo,label,sandbox\_signatures & 16,315 & 4,136 & 20,451 \\
+malware\_family & 59 & 17 & 76 \\
\hline
\textbf{Total Number of Files} & \textbf{505,407} & \textbf{126,352} & \textbf{631,759} \\
\hline
\end{tabular}
\vspace{1em}
\caption{Distribution of files within our real-world dataset by richness of metadata.}
\label{tab:dataset_metadata}
\end{table}

\subsection{Synthetic Data}

Additionally, we developed a method to generate metadata-rich training samples synthetically, using LLM-generated summaries as the required natural language description%
. We compiled C files from 44 open-source projects listed in Appendix Table \ref{tab:compilation_chunks} (see also \cite{jin2023}) into both 32-bit and 64-bit ELF files using \verb|gcc|. We used the \verb|--gdb3| option to generate debug symbols during compilation. In a second step, the \verb|gdb| debugger used these to generate views of a binary file in which each line of code is associated with its corresponding assembly and binary code representation (see Appendix Figure \ref{fig:comp_chunk}). We call a combination of C code line and corresponding compiled binary code a compilation chunk. We obtained 5,873,592 compilation chunks, totaling 213 MB. 3,010,099 of these were discarded as they only contained a single character of C code (e.g., a closing bracket), leaving 2,863,493 usable compilation chunks.

Since these are generated from a single line of code, they only refer to atomic behaviors. To obtain more complex behavioural signatures, we concatenate compilation chunks at the lowest level of nesting within their corresponding C code. This yields 294,093 aggregated chunks, which were added to the compilation chunk dataset, yielding a total of 3,157,586 chunks.

Finally, we used an LLM (Llama-3.1-Nemotron-70B-Instruct-HF) to obtain the natural language explanation of the action(s) in each chunk. Specifically, we fed each C code line/concatenation of lines into the LLM using the prompt in Appendix Listing \ref{lst:comp_prompt}. The output was attached to the chunk's binary representation and used for LBM training in a random 80/20 train/test split.

\begin{figure}[h]
  \centering
  \begin{minipage}[c]{0.6\textwidth}
    \centering
    \includegraphics[width=\linewidth]{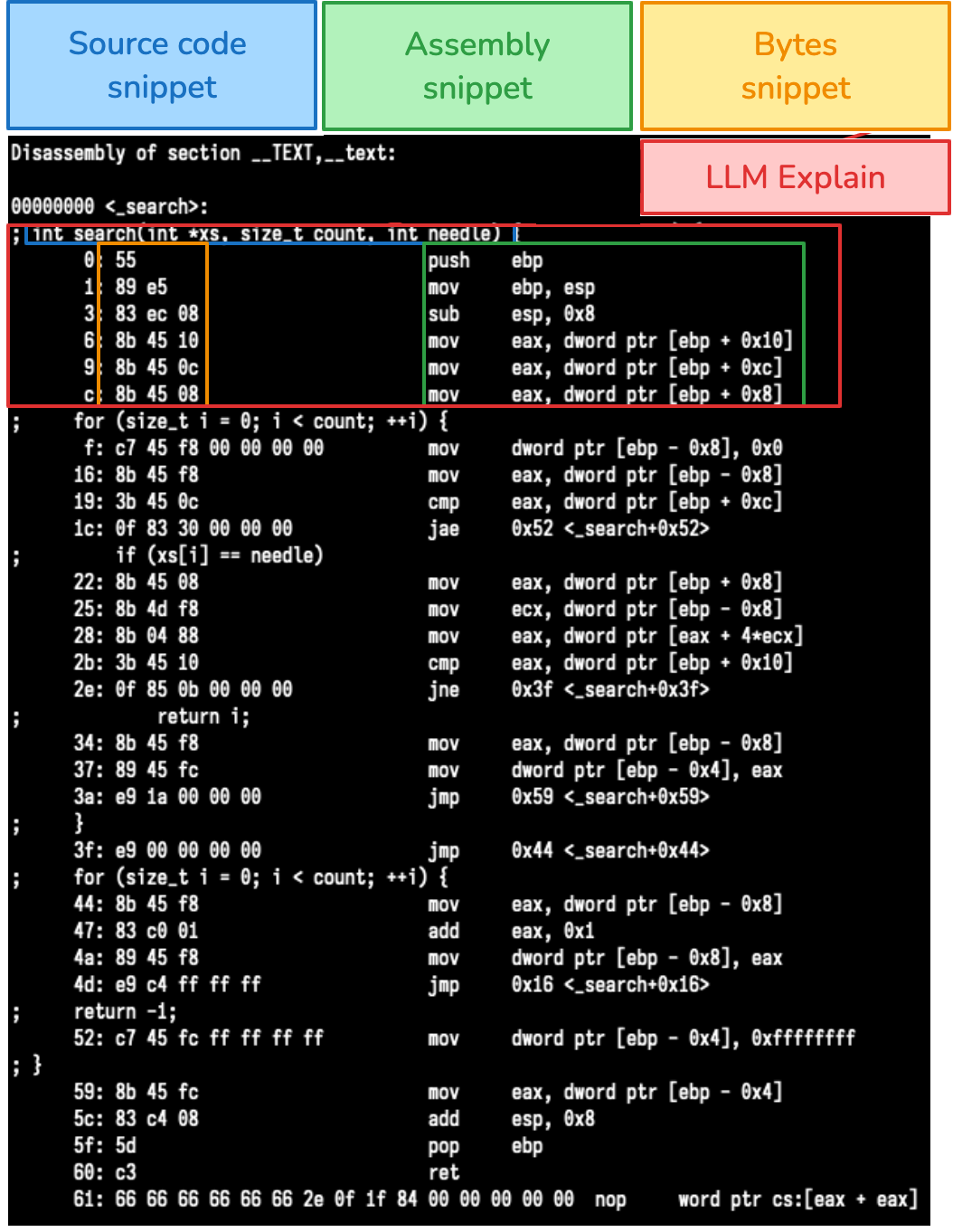}
  \end{minipage}
  \hfill
  \begin{minipage}[c]{0.38\textwidth}
    \centering
    \includegraphics[width=\linewidth]{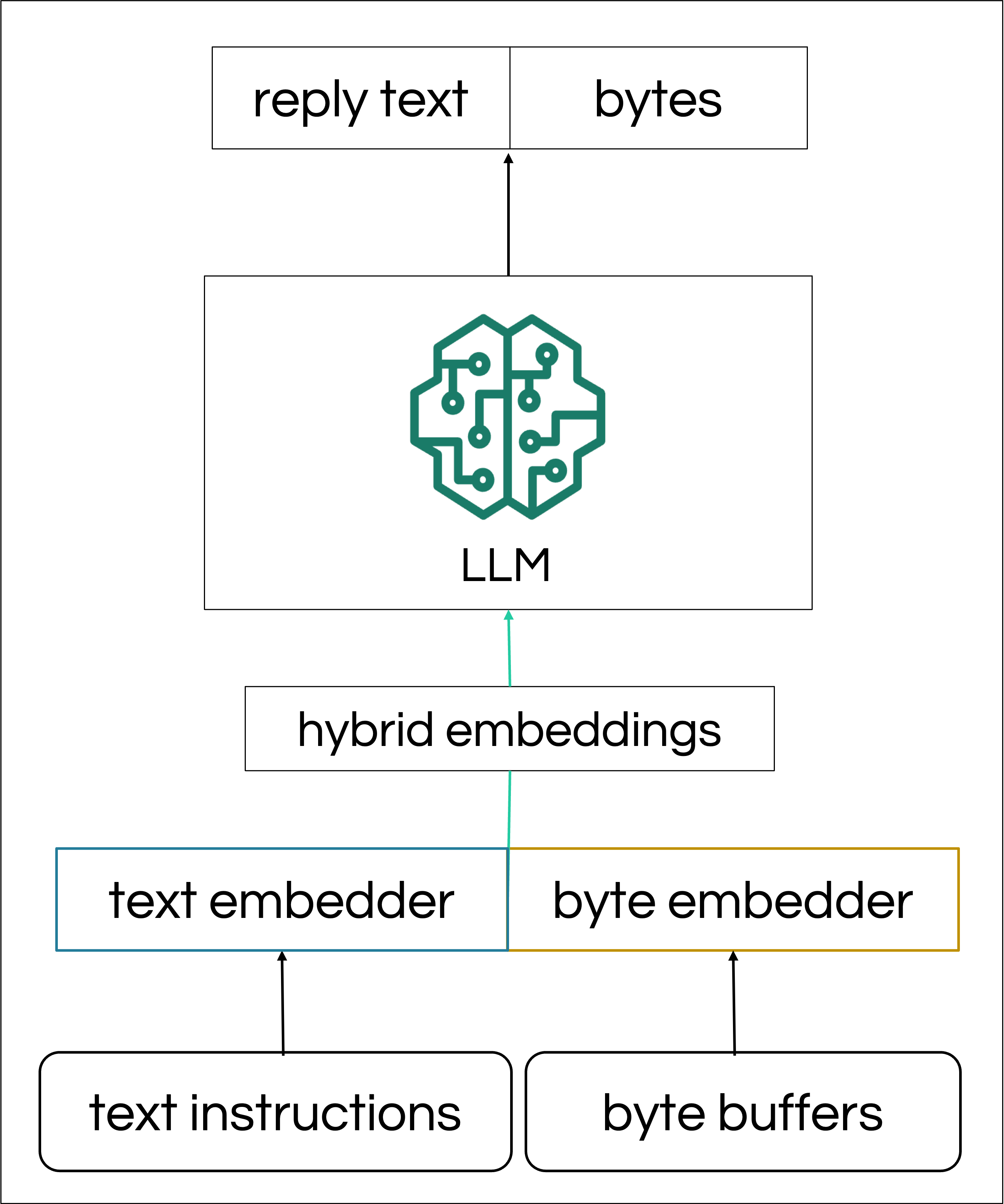}
  \end{minipage}
  \cprotect\caption{\textbf{Left:} Example output of the \verb|objdump| tool, displaying binary (orange) and assembly code (green) representations of a given line of C code (blue). These together form a compilation chunk (red), which an LLM can explain to generate synthetic data. \textbf{Right:} Hybrid tokenizer architecture.}
  \label{fig:comp_chunk}
\end{figure}

\section{Methodology}
\label{sec:Methodology}
\subsection{Model Architecture}
We use a vocabulary expansion technique \cite{kim2024} to extend the vocabulary of an off-the-shelf LLM with tokens learned from binary data, followed by continual pre-training on a combination of bytes and natural language (English text). Our goal is for the LLM to learn the structure and semantics of byte code while preserving its existing understanding of natural language. These new tokens, \textit{byte} tokens, are learned by applying the Byte Pair Encoding (BPE) algorithm to an underlying training dataset of raw binary data, in a way that represents a suitable compromise between information retention and information compression (see Appendix Figure \ref{fig:comp_rate}). These tokens, disjoint from the base model's text tokens, are passed to the LLM via a separate embedder. For training and inference, \textit{byte} embeddings are concatenated with the text embeddings and delineated by special \verb|<bytes_start>| and \verb|<bytes_end>| tokens. Together, the byte and text modalities form a hybrid embedding (see Figure \ref{fig:comp_chunk}). We initialize the newly added byte token embeddings by tokenizing each byte token with the base model's tokenizer and averaging the resulting sub-token embeddings.

\subsection{Training}
We implemented our solution using the huggingface framework in \verb|bf16| precision and its default AdamW optimiser. The training process is structured in two parts: next-byte prediction and instruction-based finetuning. 

\textbf{Next-Byte Prediction.} To ingest a large amount of byte data, we first use an unsupervised next-byte prediction approach using the newly introduced byte tokens. A cross-entropy distance between the predicted next byte(s) and the true next byte(s) based on an input data set is used to optimize both the LLM’s and the byte embedder’s internal parameters. 
To avoid catastrophic forgetting of the previously learned text embeddings, we train on natural language data from the Magpie-Pro-300K-Filtered \cite{xu2024magpie,magpie300k} and the Magicoder-Evol-Instruct-110K \cite{magicoder} datasets mixed together with byte data. %

\textbf{Instruction-Based Finetuning.} In order to connect the model’s newly gained, implicit knowledge about byte information, we perform a fine-tuning of a subset of its parameters based on instruction-response pairs. Instruction and response data both consist of corresponding text and byte elements, where the text element describes the byte element in natural language, incorporating expert knowledge about the byte buffer. This way, we introduce domain knowledge and terminology into the model without having to manually generate this information at the vast scale required to train the LLM from scratch on a new set of tokens.

\textbf{Long-Context Training.} Given that binaries can be excessively long ($\geq$1GB), we use several optimizations for long-context training, including Fully Sharded Data Parallel (FSDP \cite{zhao2023pytorchfsdpexperiencesscaling}) for multi-node multi-GPU training and DeepSpeed's sequence parallelism algorithm \cite{jacobs2023deepspeed}, which splits sequences across multiple GPUs. During the forward pass, the attention module input on each GPU has dimensions (N/SP, H, D), where SP denotes the sequence parallelism degree, H the number of attention heads, and D the hidden dimension. An all-to-all gather operation transforms this to (N, H/SP, D), allowing each GPU to compute attention for its subset of heads using optimized APIs like Flash Attention \cite{dao2023flashattention2} or cuDNN \cite{Gopal2025NVIDIA}%
, before redistributing back to the original layout. 
For memory-efficient loss computation with long contexts, we employ cut cross entropy \cite{wijmans2025} to avoid materializing all logits.

\section{Evaluation}
\label{sec:Evaluation}
\subsection{Byte Tokenizer}
To enrich the base LLM vocabulary with custom byte tokens, we trained a BPE algorithm on approximately 10GB of raw binary data, creating a vocabulary of 5,120 \textit{byte} tokens. The training dataset was balanced equally between malicious and benign files across multiple architectures (PE, ELF, MACHO, both 32 and 64-bit). The final BPE model achieves an average compression ratio (CR, defined as tokenized size divided by original size) of 0.51 on benign files and 0.57 on malicious files. Our ablation studies revealed that training exclusively on malicious files marginally increases the mean CR on benign files (0.51 → 0.55) while slightly decreasing it on malicious files (0.57 → 0.55). Architecture-specific analysis shows that malicious ARM files compress better (0.47-0.5 CR) compared to clean ARM files ($>$ 0.6 CR), while x86 clean files achieve better compression when trained solely on x86 malicious files (0.48 CR versus 0.51 CR with mixed training). These findings are relevant as the compression rate could also serve as a feature for downstream classification tasks.
Token analysis also reveals patterns including null sequences (\texttt{00, 0000}) and assembly instruction tokens such as "\texttt{mov <register>, <register>}" and "\texttt{push <ebp> / mov <ebp>, <esp>}", demonstrating the ability of the tokenizer to successfully capture byte code semantics.

\subsection{Evaluation Setup}
We evaluate the models using (i) weighted accuracy for multiclass classification tasks, including fileinfo classification, malware family classification, and benign/malicious file classification, and (ii) accuracy for opcode outlier detection. For (ii), the model is given a set of five opcodes (instructions), each in its corresponding byte form. Four of these belong to the same semantic class (out of stack movement, binary shifting, comparison, jump, %
etc., the classes are inspired from Table 8 from \cite{OutlierDataset}), and one is an outlier that belongs to another class. We then calculate the model's accuracy when identifying the outlier among these five instructions. This is done by retrieving learned embeddings for each instruction and identifying their outlier based on the average distance to the remaining embedded opcodes.

We further propose (iii) an evaluation task that measures the model’s surprise (perplexity) on so-called ‘chaotic’ bytes. To this end, we randomly generate invalid machine code that would not be observed in the wild. Two methods are used for this. One method consists of disassembling the binary, swapping each instruction with the next with probability $p$, and then reassembling the resulting machine code to obtain semantically invalid machine code. %
Another method is to randomize (automatically or manually, using the instruction encoding table) the specific x86 opcode byte or the ModR/M bytes in instructions, thereby yielding syntactically invalid opcodes.
To assess whether the model captures these invalidated binaries, we compute its perplexity $PP_{\mathrm{clean/chaotic}}$ on the altered and unaltered binaries, respectively, and then compute the Kullback-Leibler (KL) divergence $D_{\mathrm{KL}}$ between the resulting distributions across all test-set binaries. The greater the KL divergence, the more “different” the model considers those two datasets, giving us a valuable quantifier for the degree to which that particular model has learned about byte semantics. %
\begin{figure}
    \centering
    \includegraphics[width=\linewidth]{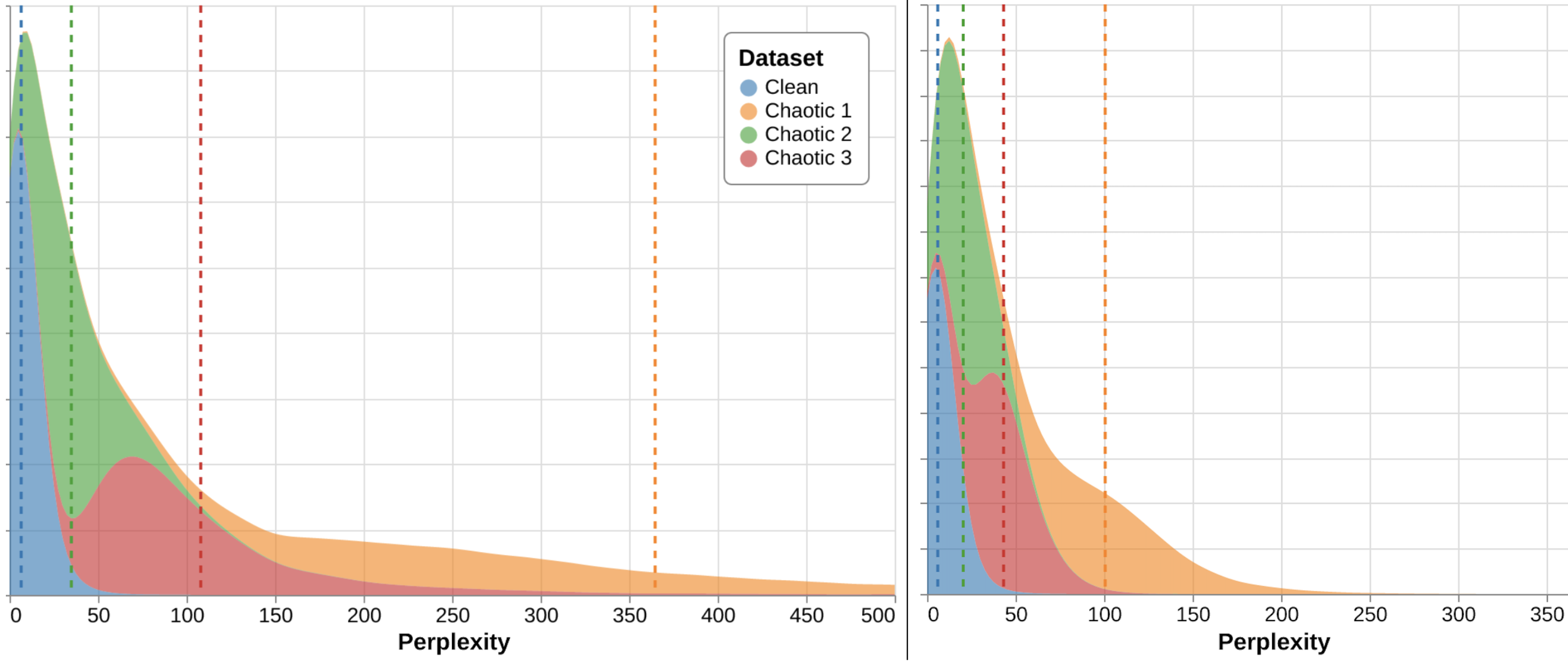}
    \caption{Llama-3.1-8B (\textbf{left}) and Mistral-7B (\textbf{right}) perplexity distributions over the malformed byte sequence datasets, where $\mathrm{Clean}$ refers to the unaltered, semantically valid byte dataset. These models have been trained using both our real-world and synthetic dataset.}
    \label{fig:kl_div_chaotic}
\end{figure}

\subsection{Results}
We trained two models, one using a Llama-3.1-8B base model and one using a Mistral-7B-Instruct-v0.2 base model, with sequence lengths of 32k and 128k tokens, respectively. For additional context in LLM selection in a corporate environment, see \autoref{sec:llm_selection}. 
Both were trained until the training loss converged. The Llama-3.1-8B based model was trained for 2.0 epochs, having ingested 402.7bn tokens and with final cross-entropy loss of 0.8935. The Mistral-7B-Instruct-v0.2 based one was trained for 0.4 epochs, having ingested 78.9bn tokens and with a final cross-entropy loss of 1.0094. %

\begin{table}[!h]
\centering
\begin{tabularx}{\columnwidth}{|X|}
\hline
\small{\textbf{Query:}} \compactcode{Can you provide some basic information about this file? }\newline 
\small{\textbf{Reply:}} \compactcode{The file under analysis has a pe, 32 bits, x86 arch (i386 machine) format structure.} \\
\hline
\small{\textbf{Query:}} \compactcode{Which malware family is this?}\newline 
\small{\textbf{Reply:}} \compactcode{The file under analysis is deemed malicious. I have identified connections to the agenttesla malware family, which is characterized as a credentialharvesting threat.} \\
\hline
\small{\textbf{Query:}} \compactcode{What behavioral characteristics can you identify from this byte buffer?}\newline 
\small{\textbf{Reply:}} \compactcode{The following behaviors are linked to this file: Creates guarded memory regions (anti-debugging trick to avoid memory dumping); Queries kernel debugger information; Queries volume information; Connects to LPC ports; Opens the Kernel Security Device Driver (KsecDD) of Windows; Touches files in the Windows directory; Installs hooks/patches the running process; Loads the .NET runtime environment; Reads the cryptographic machine GUID; Reads information about supported languages; Reads the active computer name; Overview of unique CLSIDs touched in registry; Matched Compiler/Packer signature; Contains ability to write to a remote process; Found potential URL in binary/memory; Contains PDB pathways; Found potential IP address in binary/memory; Input file contains API references not part of its Import Address Table (IAT).} \\
\hline
\small{\textbf{Query:}} \compactcode{Is this byte buffer safe or malicious?}\newline  
\small{\textbf{Reply:}} \compactcode{The file under examination is classified as malicious. It is related to the salityv4 malware family, characterized as a botnet threat.} \\
\hline
\small{\textbf{Query:}} \compactcode{Which malware family is this?}\newline  
\small{\textbf{Reply:}} \compactcode{The file under analysis is deemed malicious. I have identified connections to the xorddos malware family, which is characterized as a ddos threat.} \\
\hline
\end{tabularx}
\vspace{1em}
\caption{Sample outputs of the Llama-3.1-8B model (128k tokens context length). File sizes range from 44KB to 146KB. Replies have been verified as correct by our threat researchers. See Appendix Table \ref{tab:model_outputs_comparison} for a comparison with off-the-shelf models.}%
\label{tab:model_outputs}
\end{table}

Sample predictions by the trained Llama-3.1-8B based model are shown in Table \ref{tab:model_outputs}.
Table \ref{tab:radar_metrics} shows the models' performance on the metrics defined above. We observe that the three classification tasks differ in their difficulty for the models, with fileinfo classification accuracies ranging from 89\% to 98\%, whereas malicious/benign label classification accuracies range from 71\% to 89\%. Augmenting the training set with synthetic data consistently improves classification accuracy, underscoring the importance of this modality. The Llama-3.1-8B based model outperforms the Mistral-7B based model on the harder malware family and label classification tasks, likely due to its increased context length. 
The same goes for the perplexity-based benchmark, where the Llama-3.1-8B model's perplexity distributions for valid and invalid byte-code datasets are furthest apart as measured by $D_{\mathrm{KL}}(PP_{\mathrm{clean}} || PP_{\mathrm{Chaotic3}})$.
For outlier detection tasks, the reverse is true: low-level understanding of bytecode semantics is easier to achieve with shorter sequence lengths. 

\begin{table}[!h]
\centering
\begin{tabular}{@{}lcccccc@{}}
\toprule
 &
  \multicolumn{3}{c}{\textbf{\begin{tabular}[c]{@{}c@{}}Mistral-7B \\ 32K context\end{tabular}}} &
  \multicolumn{3}{c}{\textbf{\begin{tabular}[c]{@{}c@{}}Llama-3.1-8B \\ 128K context\end{tabular}}} \\ \cmidrule(l){2-4} \cmidrule(l){5-7}
\textbf{Task}                   & \textbf{Base} & \textbf{RW} & \textbf{+ Syn} & \textbf{Base} & \textbf{RW} & \textbf{+ Syn} \\ \midrule
Fileinfo Classification         & 6\%           & 89\%        & \textbf{98\%}  & 3\%           & 96\%        & 96\%           \\
Label Classification (F1 Score) & 22\%          & 81\%        & 82\%           & 8\%           & 71\%        & \textbf{89\%}  \\
Malware Family Classification   & 0\%           & 39\%        & 62\%           & 0\%           & 60\%        & \textbf{69\%}  \\ \midrule
Opcode Outlier Detection        &  30\%           & 36\%        & \textbf{41\%}  & 30\%            & 27\%        & 28\%           \\
KL div.\ between Perplexity dist& 0.27          & 4.37        & 8.11           & 0.40          & 10.23       & \textbf{12.49} \\ \bottomrule
\end{tabular}
\vspace{1em}
\caption{Weighted accuracy scores of several model checkpoints, which have been trained on subsets of our training data (synthetic data ablation study). Our outlier detection benchmarks are a choose-one-of-four task, in which a random guess yields a 25\% score. The KL divergence-based metric measures how well the models separate valid from invalid byte-codes based on their perplexity distributions (larger values are better). We observe that models trained on a combination of synthetic (Syn) and real-world (RW) data consistently outperform those trained only on RW data.}
\label{tab:radar_metrics}
\end{table}

Table \ref{tab:frontier_metrics} shows the performance of three frontier base models on the metrics defined above. We see that our solution surpasses all of them in terms of classification accuracy.

As a further base line, we fine-tuned 115M parameters of the Llama-3.1-8B base model on 60B train set tokens via a LoRA adapter. Table \ref{tab:frontier_metrics} shows that this yields a near-zero malware classification performance (2.5\%), a file info classification accuracy of 38.9\% and a worse than random accuracy on label classification (36.8\%). This supports our hypothesis that fine-tuning over a misaligned tokenization scheme cannot substitute for our model's full approach of vocabulary expansion, bespoke real-world and synthetic dataset curation, and dedicated byte tokenization.

\begin{table}
\centering
\begin{tabular}{@{}lccc|c@{}}
\toprule
 & \multicolumn{3}{c|}{\textbf{Off-the-shelf}} & \textbf{PEFT} \\
\cmidrule(lr){2-4} \cmidrule(l){5-5}
\textbf{Task} &
  \textbf{\begin{tabular}[c]{@{}c@{}}Nemotron 3-\\Super 120B\end{tabular}} &
  \textbf{\begin{tabular}[c]{@{}c@{}}Claude-4.5\\Sonnet\end{tabular}} &
  \textbf{\begin{tabular}[c]{@{}c@{}}Llama-4\\Maverick-17B\end{tabular}} &
  \textbf{\begin{tabular}[c]{@{}c@{}}Llama-3.1-8B\end{tabular}} \\ \midrule
Fileinfo Classification & 81\% & 74\% & 77\% & 39\% \\
Label Class. (F1 Score) & 63\% & 77\% & 69\% & 37\% \\
Malware Family Class.   & 49\% & 50\% & 48\% & 3\%  \\ \bottomrule
\end{tabular}
\vspace{1em}
\caption{Baseline comparison experiments. \textbf{Left}: Weighted accuracy scores of several frontier models evaluated on our classification benchmarks (see Table \ref{tab:radar_metrics}). These were generated by randomly selecting 400 test set samples, split evenly between malware and benign files, and submitting them to each model with a task-specific prompt. For details of the evaluation see \autoref{sec:frontier_eval}. \textbf{Right}: Weighted accuracy scores for Llama-3.1-8B fine-tuned with PEFT (Parameter-Efficient Fine-Tuning) on 115M parameters, evaluated on the entire test set. Performance is significantly lower compared to full pre-training (\autoref{tab:radar_metrics}), particularly for malware family classification.}
\label{tab:frontier_metrics}
\end{table}

\begin{figure}[h]
\centering
  \includegraphics[width=0.7\columnwidth]{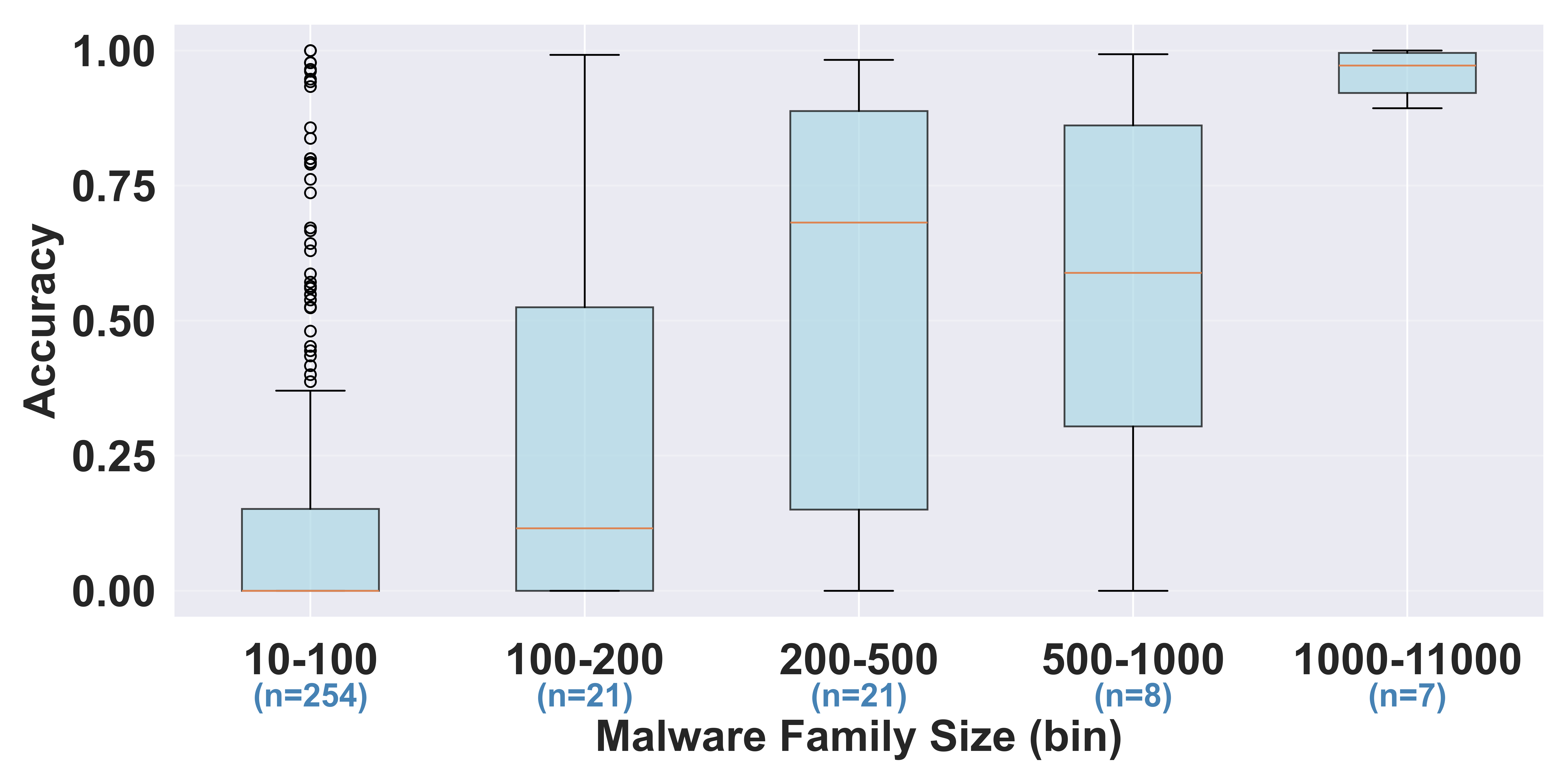}
  \caption{Malware family classification accuracy per size bin. Each bin is annotated with the number of malware families of that corresponding size.
  }
  \label{fig:malwarefam_bins}
\end{figure}

The hardest task of malware family classification shows a distinct dependence on the size of the respective malware family within our training set. Figure \ref{fig:malwarefam_bins} shows that malware families with more members present in our training set tend to achieve higher accuracy values during testing, making the case that at least a few hundred members of a given malware family are required to teach these models to accurately discern it from other families.

\textbf{Long-Context Training.} Since our multi-GPU training runs can take several days to complete, and dozens of iterations are required for experimentation with each base model, efficient use of GPU resources is imperative to achieve tight feedback loops and make our setup economically feasible.
We therefore measure training performance using Hardware FLOP Utilization (HFU) as the estimate of the ratio of FLOPs observed on a given GPU to its theoretical peak FLOPs, and Model FLOP Utilization (MFU \cite{mosaicml}) as the fraction of the GPU’s theoretical peak FLOPs that is required for computational steps during training (forward/backward passes). The latter is less dependent on implementation details and therefore more generalizable across different setups.

\begin{figure}[!h]
  \centering
  \includegraphics[width=0.75\columnwidth]{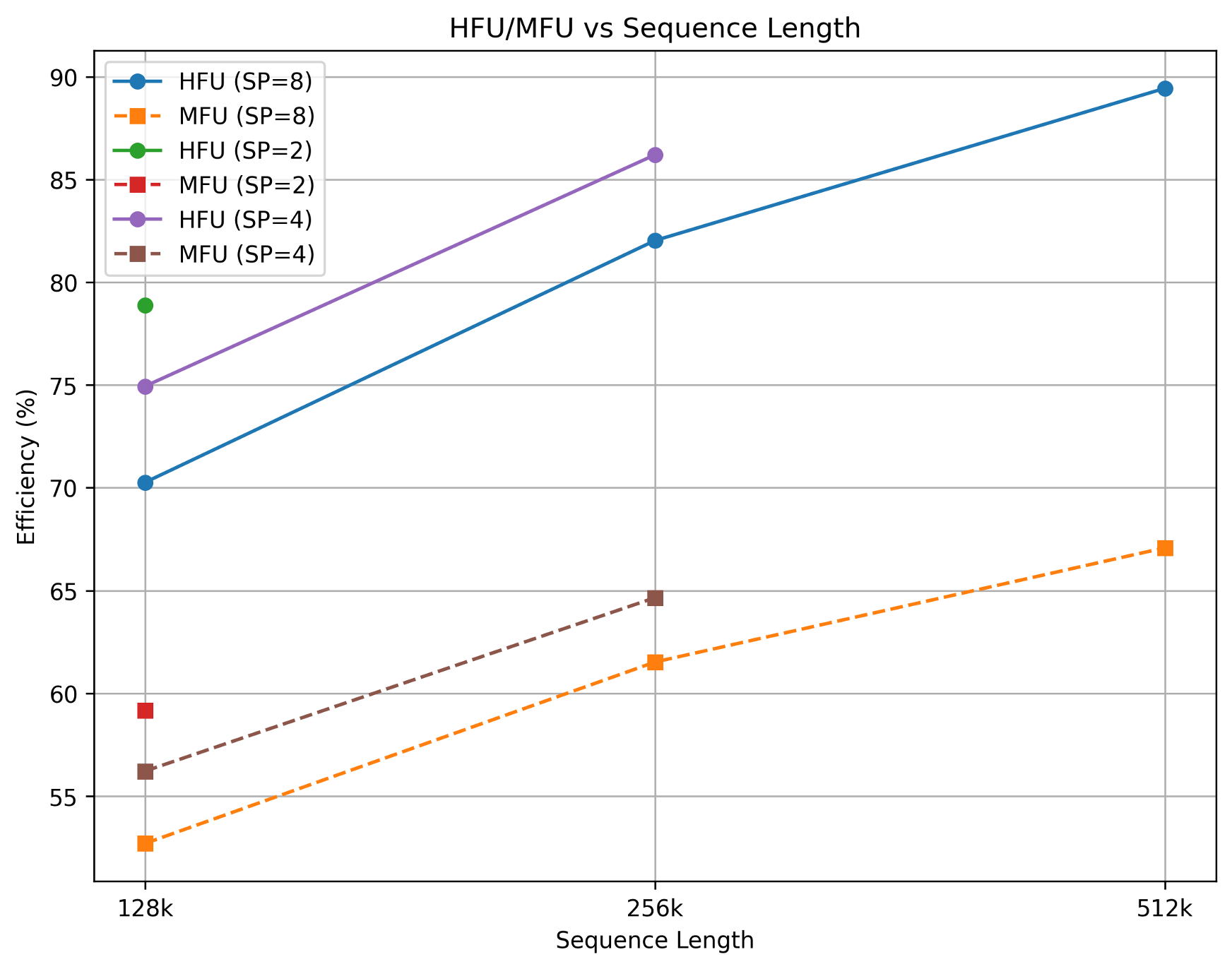}
  \caption{Llama 3.1-8B Hardware FLOPs (HFU, solid lines) and Model FLOPs (MFU, dashed lines) in dependence of sequence length, for several sequence parallelism (SP) values. The model was run with a batch size of 1, i.e., one data point was split across the number of GPUs indicated by the SP value. Larger SP values are required for larger sequence lengths under GPU memory constraints.}
  \label{fig:HFU_vs_seqlen}
\end{figure}

Figure \ref{fig:HFU_vs_seqlen} shows that a larger sequence length results in higher efficiency as measured by these two quantifiers, showing the reduced relative impact of sequence length-independent computations. For the Llama-3.1-8B model, our training routine achieves HFU values between 70\% and 90\%, indicating a GPU utilization close to its theoretical maximum, with MFU values between 50\% and 70\%, indicating that there is considerable computational overhead that does not contribute to useful model computation (e.g.\ communicational or caching overhead).

\section{Limitations}
\label{sec:limitations}
Our current evaluation setup covers a set of tasks with an exact ground truth that can be evaluated fairly and safely at scale. Evaluating the model on complex reasoning tasks  (e.g., vulnerability attribution, behavior prediction) is a valuable direction, but high quality data is scarce and costly to generate. The current classification tasks represent the highest priority identified in our survey conducted among malware researchers.

\section{Conclusion}
\label{sec:Conclusion}
Our LLM-based solution to assist threat analysts with malware analysis combines byte-vocabulary expansion %
with a large curated repository of domain-specific training data, advances in large-scale training and thorough benchmarking of the models' learned concept of byte semantics.
Trade-offs of our solution include its upper limit of 256KB per chunk due to the GPU memory footprint during training.
Giving users access to larger files is a focal point of ongoing development before taking this emerging solution from limited deployment to general availability.

Learnings about model architecture, training process and data composition have the potential for considerable positive societal impact, since their application to malware analysis and detection can make it harder for adversaries to infiltrate computer systems. 
As always in this field, improved algorithms can also lead to increasingly sophisticated detection evasion. A potentially negative societal impact lies in the over-reliance on automatic analysis systems without appropriate human supervision.

\newpage
\bibliographystyle{unsrtnat}
\bibliography{bibliography}

\begin{thebibliography}{23}
\providecommand{\natexlab}[1]{#1}
\providecommand{\url}[1]{\texttt{#1}}
\expandafter\ifx\csname urlstyle\endcsname\relax
  \providecommand{\doi}[1]{doi: #1}\else
  \providecommand{\doi}{doi: \begingroup \urlstyle{rm}\Url}\fi

\bibitem[CrowdStrike()]{csthreat25}
CrowdStrike.
\newblock Crowdstrike 2025 global threat report.
\newblock \url{https://www.crowdstrike.com/explore/2025-global-threat-report-en-gb}.

\bibitem[He et~al.(2026)He, Fox, Miculicich, Friedli, Fabian, Gokturk, Tang, Lee, Pfister, and Le]{he2026}
Pengfei He, Ash Fox, Lesly Miculicich, Stefan Friedli, Daniel Fabian, Burak Gokturk, Jiliang Tang, Chen-Yu Lee, Tomas Pfister, and Long~T. Le.
\newblock Co-redteam: Orchestrated security discovery and exploitation with llm agents, 2026.
\newblock URL \url{https://arxiv.org/abs/2602.02164}.

\bibitem[Yu et~al.(2023)Yu, Simig, Flaherty, Aghajanyan, Zettlemoyer, and Lewis]{yu2023}
Lili Yu, Dániel Simig, Colin Flaherty, Armen Aghajanyan, Luke Zettlemoyer, and Mike Lewis.
\newblock Megabyte: Predicting million-byte sequences with multiscale transformers, 2023.
\newblock URL \url{https://arxiv.org/abs/2305.07185}.

\bibitem[Quintero()]{gemini24}
Bernardo Quintero.
\newblock Gemini for malware analysis.
\newblock \url{https://cloud.google.com/blog/topics/threat-intelligence/gemini-for-malware-analysis}.

\bibitem[Wu et~al.(2024)Wu, Tan, Wang, Wang, Li, and Sun]{wu2024}
Shangda Wu, Xu~Tan, Zili Wang, Rui Wang, Xiaobing Li, and Maosong Sun.
\newblock Beyond language models: Byte models are digital world simulators, 2024.
\newblock URL \url{https://arxiv.org/abs/2402.19155}.

\bibitem[Pagnoni et~al.(2024)Pagnoni, Pasunuru, Rodriguez, Nguyen, Muller, Li, Zhou, Yu, Weston, Zettlemoyer, Ghosh, Lewis, Holtzman, and Iyer]{pagnoni2024}
Artidoro Pagnoni, Ram Pasunuru, Pedro Rodriguez, John Nguyen, Benjamin Muller, Margaret Li, Chunting Zhou, Lili Yu, Jason Weston, Luke Zettlemoyer, Gargi Ghosh, Mike Lewis, Ari Holtzman, and Srinivasan Iyer.
\newblock Byte latent transformer: Patches scale better than tokens, 2024.
\newblock URL \url{https://arxiv.org/abs/2412.09871}.

\bibitem[aii()]{aiind24}
Stanford ai index.
\newblock \url{https://aiindex.stanford.edu/wp-content/uploads/2024/05/HAI_AI-Index-Report-2024.pdf}.

\bibitem[Kim et~al.(2024)Kim, Choi, and Jeong]{kim2024}
Seungduk Kim, Seungtaek Choi, and Myeongho Jeong.
\newblock Efficient and effective vocabulary expansion towards multilingual large language models, 2024.
\newblock URL \url{https://arxiv.org/abs/2402.14714}.

\bibitem[Team(2026)]{radare2}
Radare2 Team.
\newblock Radare2 github repository.
\newblock \url{https://github.com/radare/radare2}, 2026.

\bibitem[Jin et~al.(2023)Jin, Larson, Yang, and Lin]{jin2023}
Xin Jin, Jonathan Larson, Weiwei Yang, and Zhiqiang Lin.
\newblock Binary code summarization: Benchmarking chatgpt/gpt-4 and other large language models, 2023.
\newblock URL \url{https://arxiv.org/abs/2312.09601}.

\bibitem[Xu et~al.(2024)Xu, Jiang, Niu, Deng, Poovendran, Choi, and Lin]{xu2024magpie}
Zhangchen Xu, Fengqing Jiang, Luyao Niu, Yuntian Deng, Radha Poovendran, Yejin Choi, and Bill~Yuchen Lin.
\newblock Magpie: Alignment data synthesis from scratch by prompting aligned llms with nothing, 2024.
\newblock URL \url{https://arxiv.org/abs/2406.08464}.

\bibitem[mag({\natexlab{a}})]{magpie300k}
Magpie-pro-300k-filtered.
\newblock \url{https://huggingface.co/datasets/Magpie-Align/Magpie-Pro-300K-Filtered}, {\natexlab{a}}.

\bibitem[mag({\natexlab{b}})]{magicoder}
Magicoder-evol-instruct-110k.
\newblock \url{https://huggingface.co/datasets/ise-uiuc/Magicoder-Evol-Instruct-110K}, {\natexlab{b}}.

\bibitem[Zhao et~al.(2023)Zhao, Gu, Varma, Luo, Huang, Xu, Wright, Shojanazeri, Ott, Shleifer, Desmaison, Balioglu, Damania, Nguyen, Chauhan, Hao, Mathews, and Li]{zhao2023pytorchfsdpexperiencesscaling}
Yanli Zhao, Andrew Gu, Rohan Varma, Liang Luo, Chien-Chin Huang, Min Xu, Less Wright, Hamid Shojanazeri, Myle Ott, Sam Shleifer, Alban Desmaison, Can Balioglu, Pritam Damania, Bernard Nguyen, Geeta Chauhan, Yuchen Hao, Ajit Mathews, and Shen Li.
\newblock Pytorch fsdp: Experiences on scaling fully sharded data parallel, 2023.
\newblock URL \url{https://arxiv.org/abs/2304.11277}.

\bibitem[Jacobs et~al.(2023)Jacobs, Tanaka, Zhang, Zhang, Song, Rajbhandari, and He]{jacobs2023deepspeed}
Sam~Ade Jacobs, Masahiro Tanaka, Chengming Zhang, Minjia Zhang, Shuaiwen~Leon Song, Samyam Rajbhandari, and Yuxiong He.
\newblock Deepspeed ulysses: System optimizations for enabling training of extreme long sequence transformer models.
\newblock \emph{arXiv preprint arXiv:2309.14509}, 2023.

\bibitem[Dao(2024)]{dao2023flashattention2}
Tri Dao.
\newblock Flash{A}ttention-2: Faster attention with better parallelism and work partitioning.
\newblock In \emph{International Conference on Learning Representations (ICLR)}, 2024.

\bibitem[Gopal et~al.(2025)Gopal, Macchi, Baker, Knight, Zhang, Valgur, Watanabe, Moon, Agarwalla, and {swimvtec}]{Gopal2025NVIDIA}
Anerudhan Gopal, Emilien Macchi, Connor Baker, James~Y Knight, Jun Zhang, Martin Valgur, Takeshi Watanabe, Tim Moon, Vedaanta Agarwalla, and {swimvtec}.
\newblock Nvidia/cudnn-frontend.
\newblock https://github.com/NVIDIA/cudnn-frontend, dec 20 2025.
\newblock URL \url{https://github.com/NVIDIA/cudnn-frontend}.

\bibitem[Wijmans et~al.(2025)Wijmans, Huval, Hertzberg, Koltun, and Krähenbühl]{wijmans2025}
Erik Wijmans, Brody Huval, Alexander Hertzberg, Vladlen Koltun, and Philipp Krähenbühl.
\newblock Cut your losses in large-vocabulary language models, 2025.
\newblock URL \url{https://arxiv.org/abs/2411.09009}.

\bibitem[Li et~al.(2021)Li, Qu, and Yin]{OutlierDataset}
Xuezixiang Li, Yu~Qu, and Heng Yin.
\newblock Palmtree: Learning an assembly language model for instruction embedding.
\newblock In \emph{Proceedings of the 2021 ACM SIGSAC Conference on Computer and Communications Security}, CCS '21, page 3236–3251, New York, NY, USA, 2021. Association for Computing Machinery.
\newblock ISBN 9781450384544.
\newblock \doi{10.1145/3460120.3484587}.
\newblock URL \url{https://doi.org/10.1145/3460120.3484587}.

\bibitem[mos()]{mosaicml}
Databricks mosaic ml.
\newblock \url{https://github.com/mosaicml/llm-foundry/blob/main/scripts/train/benchmarking/README.md}.

\bibitem[Raff et~al.(2018)Raff, Barker, Sylvester, Brandon, Catanzaro, and Nicholas]{MalConv}
Edward Raff, Jon Barker, Jared Sylvester, Robert Brandon, Bryan Catanzaro, and Charles Nicholas.
\newblock Malware {Detection} by {Eating} a {Whole} {EXE}.
\newblock In \emph{{AAAI} {Workshop} on {Artificial} {Intelligence} for {Cyber} {Security}}, October 2018.
\newblock URL \url{http://arxiv.org/abs/1710.09435}.
\newblock arXiv: 1710.09435.

\bibitem[Raff et~al.(2021)Raff, Fleshman, Zak, Anderson, Filar, and McLean]{Raff2020b}
Edward Raff, William Fleshman, Richard Zak, Hyrum~S. Anderson, Bobby Filar, and Mark McLean.
\newblock Classifying {Sequences} of {Extreme} {Length} with {Constant} {Memory} {Applied} to {Malware} {Detection}.
\newblock In \emph{The {Thirty}-{Fifth} {AAAI} {Conference} on {Artificial} {Intelligence}}, 2021.
\newblock URL \url{http://arxiv.org/abs/2012.09390}.
\newblock arXiv: 2012.09390.

\bibitem[Rudd et~al.(2022)Rudd, Rahman, and Tully]{10.1145/3494110.3528242}
Ethan~M Rudd, Mohammad~Saidur Rahman, and Philip Tully.
\newblock Transformers for {End}-to-{End} {InfoSec} {Tasks}: {A} {Feasibility} {Study}.
\newblock In \emph{Proceedings of the 1st {Workshop} on {Robust} {Malware} {Analysis}}, pages 21--31, New York, NY, USA, 2022. Association for Computing Machinery.
\newblock ISBN 978-1-4503-9179-5.
\newblock \doi{10.1145/3494110.3528242}.
\newblock URL \url{https://doi.org/10.1145/3494110.3528242}.
\newblock Series Title: WoRMA '22.

\end{thebibliography}

\appendix
\thispagestyle{empty}

\onecolumn

\section{A Note On LLM Family Selection} \label{sec:llm_selection}

We explicitly note that this study has not benefitted from the abundance of resources in frontier AI labs; each of these experiments has required a considerable allocation of limited GPU resources and has created conflicts in experimental design. At each stage, we must decide between testing new augmentation and training methodologies using prior LLMs, to enable apples-to-apples comparisons, versus moving our experiments to newer LLM architectures. Predominantly, we have preferred the latter approach due to the highly atypical nature of cybersecurity data and the discrepancies between combined assembly representations and those used in more common NLP tasks in language. Indeed, many prior works have noted that the en-vogue methods for training convolutional neural networks or~\cite{MalConv, Raff2020b}, more recently, transformers ~\cite{10.1145/3494110.3528242} do not consistently generalize to cybersecurity data. 

An additional factor in LLM selection in a corporate environment is the provenance and licensing under which different LLMs are made available. For example, the Kimi-K2 license\footnote{\url{https://huggingface.co/moonshotai/Kimi-K2-Instruct/blob/main/LICENSE}} has restrictions requiring including their name, and trigger conditions in terms of revenue and number of users, which are independently problematic for us to consider both in terms of branding and potentially ``stickyness'' of such naming, but also ambiguity around specific terms in the license. For example, the ``derivative works thereof'' clause can be interpreted similarly to ``copy-left'' licenses such as the GNU Public License (GPL), and make certain evaluations problematic or require extensive legal review to even test. Compounding these concerns on top of the technical factors we wish to consider, e.g., natural sequence length available, it causes significant friction to update to more recent LLMs. 

\section{Insights from Surveying Cybersecurity Analysts}

During LBM development, we surveyed several internal cybersecurity analysts to understand how they conduct threat analysis and to identify capabilities that best serve their needs. Analysts currently rely on cached knowledge stores for prior analyses, they employ a mix of static and dynamic analysis, and utilize rule-based tools like YARA alongside similarity search and clustering techniques. Their analysis follows an iterative process: they begin with metadata inspection, then progressively zoom in by querying file characteristics such as section counts, names and entropy, imports/exports, appended data, resource sections. They frequently examine strings, disassembled code, and decompiled representations, switching between these views as needed. Primary challenges that hinder analysis include obfuscation, evasion techniques, and packed or even damaged files.

High-priority tasks that an LBM should tackle include attribution (malicious/benign/PuP), zooming-in on malware families, mapping to standard behaviors, tactics and techniques, and providing explanations (e.g., "identified as ransomware due to encryption and file-searching behaviors"). Analysts also emphasized the need for a unified Q\&A assistant that can query external tools, provide "at-a-glance" summaries, flag suspicious signals or sections for focused attention, and smooth out architecture-specific differences. Equally important is the ability to navigate seamlessly between bytes, assembly, and decompiled code to explain binary intent at multiple levels.

\section{Frontier Model Evaluation}
\label{sec:frontier_eval}
File info and label assessment were done using fuzzy matching between the model prediction and the respective ground truth label using fuzzywuzzy’s \texttt{fuzz.token\_set\_ratio()} method with a score threshold of 45\% to accommodate the natural variation in how LLMs phrase identical answers (e.g. "Windows x86 32-bit" vs "Win32" vs "32-bit Windows PE"). This threshold was deliberately set to be permissive toward the baseline models, ensuring valid answers are not rejected due to phrasing variation. The assessment of malware family classification accuracy was performed manually by a trained malware analyst. Full prompt details and evaluation scripts are available upon request.

\newpage
\section{Additional Figures and Tables}

\subsection*{Real World Dataset Composition}
\begin{table}[h]
\centering
{
\begin{tabular}{@{}lcc@{}}
\toprule
\textbf{Label} & \textbf{Count} & \textbf{Proportion} \\ \midrule
Malicious & 270406 & 0.4280 \\
Benign    & 207208 & 0.3280 \\
Adware    & 154145 & 0.2440 \\ \bottomrule
\end{tabular}
}
\vspace{1em}
\caption{Label Distribution}
\end{table}

\begin{table}[h]
\centering
{
\begin{tabular}{@{}lcc@{}}
\toprule
\textbf{Binary Type} & \textbf{Count} & \textbf{Proportion} \\ \midrule
PE    & 516359 & 0.8173 \\
ELF   &  97708 & 0.1547 \\
Mach-O &  17692 & 0.0280 \\ \bottomrule
\end{tabular}
}
\vspace{1em}
\caption{Binary Type Distribution}
\end{table}

\begin{table}[h]
\centering
{
\begin{tabular}{@{}lcc@{}}
\toprule
\textbf{Malware Family} & \textbf{Count} & \textbf{Proportion} \\ \midrule
XorDDoS         & 53595 & 0.1722 \\
Linkury         & 50492 & 0.1622 \\
Virlock         & 18259 & 0.0587 \\
CobaltStrike    & 15941 & 0.0512 \\
Mirai           & 14072 & 0.0452 \\
RedLineStealer  &  8132 & 0.0261 \\
GandCrab        &  6843 & 0.0220 \\
InstallCore     &  5470 & 0.0176 \\
BrowseFox       &  4492 & 0.0144 \\
Berbew          &  4281 & 0.0138 \\
Adposhel        &  4110 & 0.0132 \\
DotDo           &  3924 & 0.0126 \\
SalityV4        &  3901 & 0.0125 \\
Trickler        &  3845 & 0.0124 \\
AmadeyLoader    &  3841 & 0.0123 \\
AgentTesla      &  2370 & 0.0076 \\
NetInfoNabster  &  2263 & 0.0073 \\
Gator           &  2235 & 0.0072 \\
Mofksys         &  2147 & 0.0069 \\
DownloadGuide   &  1981 & 0.0064 \\
\textless{}OTHER\textgreater{} & 99114 & 0.3184 \\ \bottomrule
\end{tabular}
}
\vspace{1em}
\caption{Top-20 Malware Family Distribution}
\end{table}

\begin{table}[h]
\centering
{
\begin{tabular}{@{}lcc@{}}
\toprule
\textbf{Bits} & \textbf{Count} & \textbf{Proportion} \\ \midrule
32 & 623592 & 0.9871 \\
16 &   8167 & 0.0129 \\ \bottomrule
\end{tabular}
}
\vspace{1em}
\caption{Bit Width Distribution}
\end{table}

\begin{table}[h]
\centering
{
\begin{tabular}{@{}lcc@{}}
\toprule
\textbf{Machine} & \textbf{Count} & \textbf{Proportion} \\ \midrule
i386          & 513185 & 0.8123 \\
Intel 80386   &  73112 & 0.1157 \\
ARM           &  21010 & 0.0333 \\
386           &  10202 & 0.0161 \\
Unknown       &   3001 & 0.0048 \\
\textless{}OTHER\textgreater{} & 11249 & 0.0178 \\ \bottomrule
\end{tabular}
}
\vspace{1em}
\caption{Top-5 Machine Architecture Distribution}
\end{table}

\begin{table}[h]
\centering
{
\begin{tabular}{@{}lcc@{}}
\toprule
\textbf{Compiler} & \textbf{Count} & \textbf{Proportion} \\ \midrule
N/A     & 543119 & 0.8597 \\
GCC     &  69229 & 0.1096 \\
Clang   &  17704 & 0.0280 \\
Android &    821 & 0.0013 \\
\textless{}OTHER\textgreater{} & 886 & 0.0014 \\ \bottomrule
\end{tabular}
}
\vspace{1em}
\caption{Compiler Distribution}
\end{table}

\begin{table}[h!]
{\small
\centering
\begin{tabular}{lr|lr}
\hline
\textbf{project name} & \textbf{number of byte chunks} & \textbf{project name} & \textbf{number of byte chunks} \\
\hline
openssl & 1,555,401 & tar & 45,121 \\
gettext & 700,278 & wget & 42,010 \\
binutils\_gdb & 565,633 & nettle & 35,090 \\
curl & 358,040 & lightning & 32,842 \\
libredwg & 320,195 & findutils & 32,133 \\
freeipmi & 312,073 & adns & 29,329 \\
mailutils & 295,253 & nano & 27,033 \\
poke & 168,685 & readline & 24,172 \\
ncurses & 121,705 & less & 22,327 \\
coreutils & 118,250 & libiconv & 21,228 \\
bash & 114,707 & diffutils & 21,141 \\
guile & 98,072 & grep & 20,152 \\
dico & 97,506 & direvent & 19,302 \\
gmp & 81,876 & datamash & 17,991 \\
libmicrohttpd & 76,139 & cflow & 15,694 \\
libunistring & 71,904 & libidn2 & 14,254 \\
libpng & 67,365 & sed & 13,780 \\
gawk & 67,194 & texinfo & 9,776 \\
inetutils & 55,322 & gzip & 8,557 \\
mpfr & 55,210 & units & 7,826 \\
gama & 52,371 & gss & 7,039 \\
bison & 46,861 & libtool & 5,628 \\
\hline
\end{tabular}
\vspace{1em}
\caption{Number of byte chunks extracted from compiling open-source projects. The following compilers and settings were used: GNU gdb (Ubuntu 15.0.50.20240403-0ubuntu1) 15.0.50.20240403-git, GNU Make 4.3 Ubuntu 24.04.2 LTS (Noble Numbat), gcc (Ubuntu 13.3.0-6ubuntu2~24.04) 13.3.0. Compilation flags 32bit: \texttt{-Wno-error -m32 -ggdb3 -O0}, file: \texttt{ELF, 32bit (Intel 80386, LSB relocatable)}. Compilation flags 64bit: \texttt{-Wno-error -ggdb3 -O0}, file: \texttt{ELF, 64bit (x86-64, LSB relocatable)}
}
\label{tab:compilation_chunks}
}
\end{table}

\clearpage
\lstset{basicstyle=\ttfamily\footnotesize}
\begin{lstlisting}[
    caption=Model prompt to generate synthetic compilation data, 
    captionpos=b,
    label=lst:comp_prompt, 
    basicstyle=\ttfamily\footnotesize
]
You are a great threat analyst and expert when it comes to IT administration.

Your task is to summarise chunks of C/C++ source code (lines) and their corresponding disassembly representation, as given by GDB.
Provide a straightforward answer without prefacing it with a summary statement.
Use a generic tone in passive voice, no opinions and no recommendations.

DO NOT say anything like: "The provided code is a mix of C source code and its corresponding disassembled machine code.", just summarise the chunk.
DO NOT say anything like: "The disassembly shows...". Instead of "The disassembly shows a move operation of a 32-bit value", say "A 32-bit value is moved."
DO NOT refer to specific labels (symbols, names) from the code, focus more on the structure and general behavior. For example:
- if you see a `call address <NAME+4>` do not refer to its specific `NAME`, just say that a function is being called.
- if you see `ok = false` do not say "the code sets a boolean variable `ok` to `false`", just "a variable is being set to false", following the disassembly code.
DO NOT use extremely specific info, focus more on the disassembled code rather then on the source code.

Use source code whenever you are not confident in the meaning of the disassembly.
Try to not explain disassembly line-by-line, instead focus more on the broader behavior.

Here is the source code:

<source>
{{ code }}
</source>

Here is the assembly:
<assembly>
{{ asm }}
</assembly>

Reply:
\end{lstlisting}

\begin{figure}[!h]
  \centering
  \includegraphics[width=0.7\columnwidth]{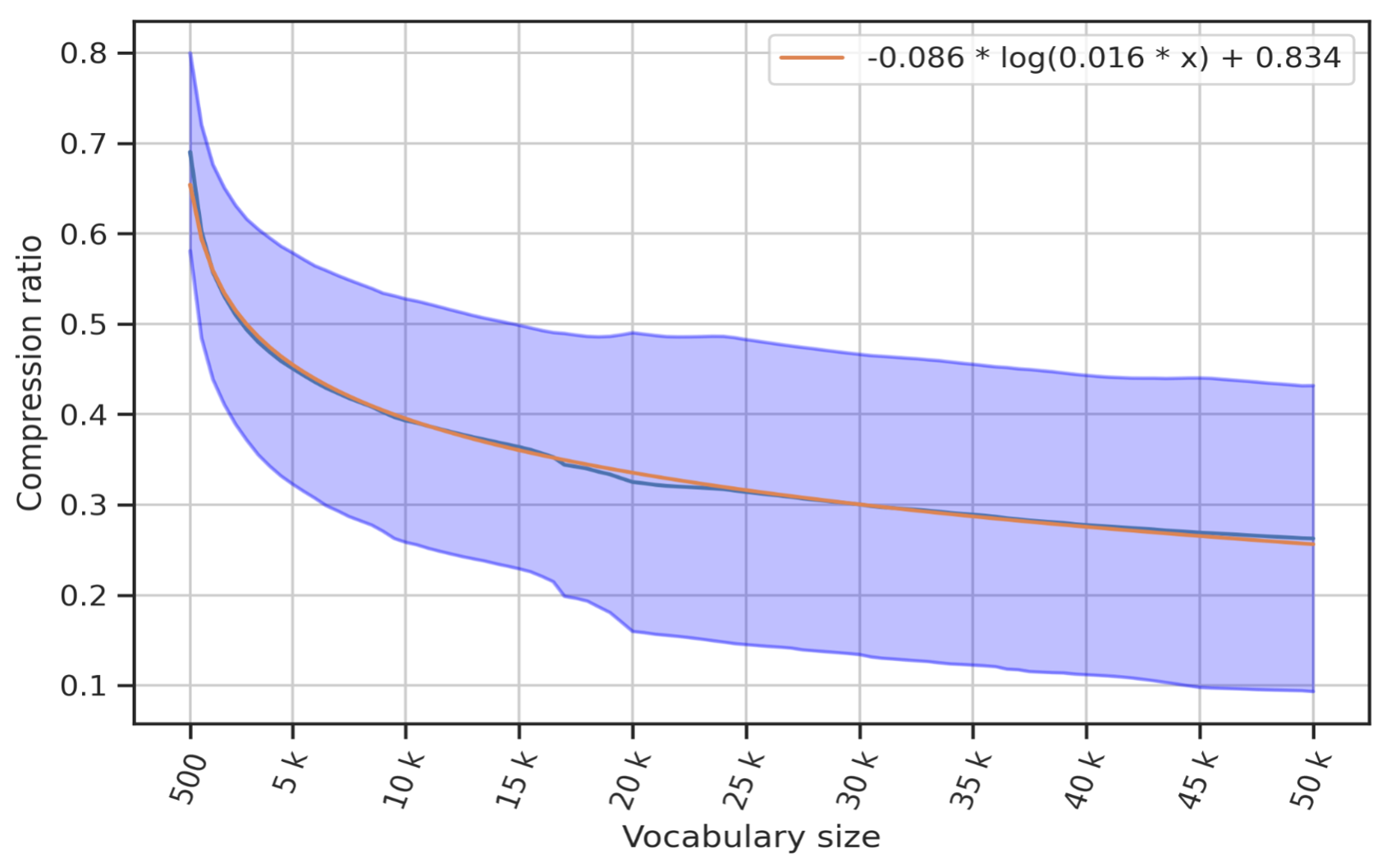}
  \caption{Tokenizer compression rate in dependence of chosen vocabulary size. The blue region represents $\mu_{CR} \pm \sigma_{CR}^2$
  }
  \label{fig:comp_rate}
\end{figure}

\begin{figure}
    \centering
    \includegraphics[width=\linewidth]{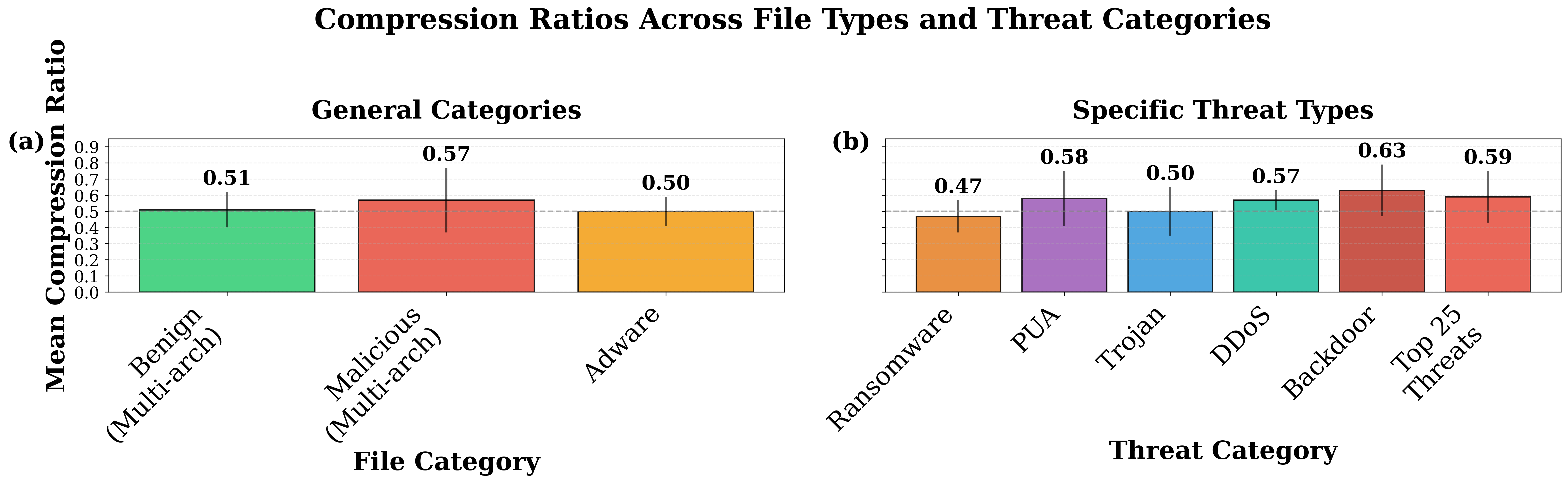}
    \caption{Mean Tokenizer Compression Ratio across file and threat categories.}
    \label{fig:cr}
\end{figure}

\begin{figure}[!h]
  \centering
  \includegraphics[width=0.7\columnwidth]{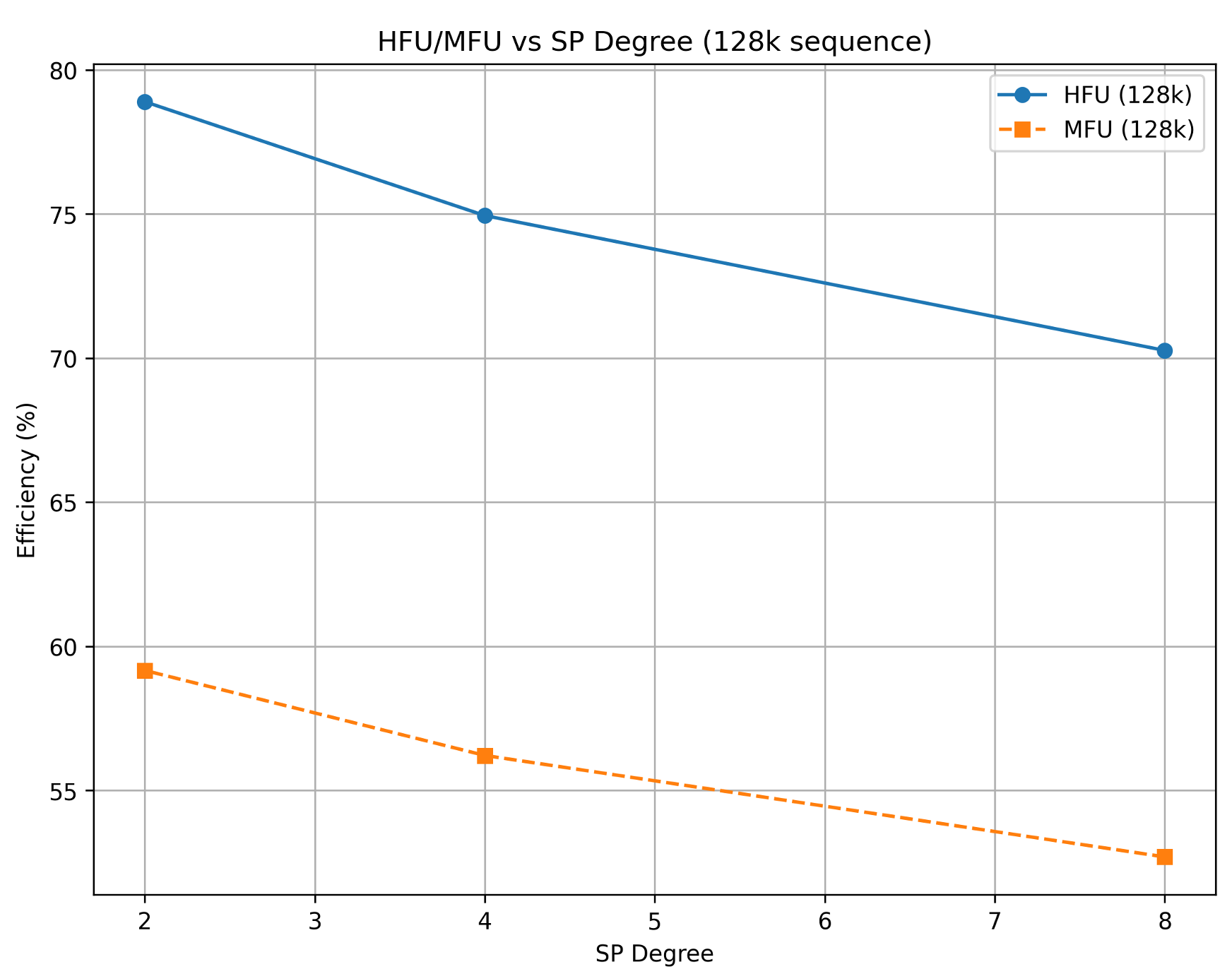}
  \caption{Hardware FLOPs usage (HFU) and Model FLOPs usage (MFU) in dependence of sequence parallelism degree (SP).}
  \label{fig:HFU_vs_SP}
\end{figure}

\newcommand{\compactcodeappendix}[1]{%
  \begin{minipage}{\linewidth}
  \setstretch{0.6} %
  \lstset{basicstyle=\ttfamily\tiny} %
  \lstinline|#1|
  \end{minipage}
}

\lstset{basicstyle=\ttfamily\tiny}
\begin{table}[h]
\centering
\begin{tabularx}{\textwidth}{|p{7em}||p{12em}|X|X|}
\hline
\textbf{Query} & \textbf{Claude Sonnet 4.5} & \textbf{Llama-3.1 8B} & \textbf{Mistral-7B} \\
\hline
\multicolumn{2}{|l|}{{\tiny sha256: \smallinline{2d18440cd33a632db4305a1cfc552facf7591668e83cf34c5851cc94afd78aa8}}}  & & \\
\compactcodeappendix{Can you provide some basic information about this file?} & \compactcodeappendix{This is a Windows PE (Portable Executable) file, specifically a .NET executable compiled for the .NET Framework 4.0. File Format: PE32 executable (console), Architecture: x86 (32-bit), Framework: .NET Framework 4.0, Assembly Name: Happy.exe (internal name: Implosions.exe), File Version: 0.0.0.0} & \compactcodeappendix{This is a binary file.} & \compactcodeappendix{This file appears to be a binary data file with a size of approximately 1.1 MB. The file starts with the byte sequence "4d 5a 90 00 03 00 00 00 04 ..."}\\
\hline
\multicolumn{2}{|l|}{{\tiny sha256: \smallinline{771ffbd2c7b81f4f4617b0dd492cdce756ed7013765c5ade8b38769add490ca2}}}  & &\\
\compactcodeappendix{Which malware family is this?} & \compactcodeappendix{This is definitively Agent Tesla, a .NET-based information stealer and remote access trojan that has been active since at least 2014.} & (empty) & \compactcodeappendix{The provided binary data appears to be a PE32 executable file with multiple sections and imports. It contains several strings that suggest it might be a variant of the (*Emotet malware family*), which is known for its modular design and ability to download and execute additional malware modules. However, without further analysis and decoding of the strings and code within the file, it is not possible to definitively determine if this is Emotet or another malware family.} \\
\hline
\multicolumn{2}{|l|}{{\tiny sha256: \smallinline{771ffbd2c7b81f4f4617b0dd492cdce756ed7013765c5ade8b38769add490ca2}}}  & & \\
\compactcodeappendix{What behavioral characteristics can you identify from this byte buffer?} & \compactcodeappendix{1. Information Stealing Capabilities (Password Recovery: Targets multiple browsers, FTP Credentials, Email Clients, Messaging Apps), 2. Keylogging \& Surveillance (Implements keyboard hook (kbHook_KeyDown, kbHook_KeyUp), Captures clipboard data, Takes screenshots (SendScreen_Tick), Webcam capture functionality (Sendwebcam_Tick)), 3. Persistence Mechanisms (Registry modifications for startup (Run key), Creates scheduled tasks, File system manipulation (copies to Java update paths), UAC bypass attempts (BypassUAC, disableuac)), 4. Anti-Analysis Features (Detects virtual machines, Checks for analysis tools, Disables security features: Task Manager (DisableTaskMgr), Command Prompt (DisableCMD), Registry Editor (DisableRegistryTools), System Restore (DisableSR)), 5. Data Exfiltration (FTP upload capabilities, SMTP email sending, HTTP POST to command \& control servers Imgur API integration for image uploads), 6. System Manipulation (Process injection capabilities, Memory manipulation, File operations, Registry key manipulation), 7. Evasion Techniques (Mutex creation for single instance, Process hollowing indicators, Encrypted data storage, Base64 encoding for data transmission)} & \compactcodeappendix{The byte buffer is a binary file. It is a Windows executable file. It is a PE32 executable (GUI) Intel 80386 32-bit. It is a Windows 32-bit EXE file.} & \compactcodeappendix{From the given byte buffer, it appears to be a binary file containing data for a PE (Portable Executable) file. The buffer starts with the signature "4d 5a 90 00 03 00 00 00 04 ...} \\
\hline
\multicolumn{2}{|l|}{{\tiny sha256: \smallinline{0cf5264f258303c5f59f66cc010c161bb49d72e3f4025a41f2d8ac87510092c7}}}  & & \\
\compactcodeappendix{Is this byte buffer safe or malicious?} & \compactcodeappendix{This is a malicious executable - specifically a (*7-Zip installer*) that has been modified or is being misrepresented. While it contains legitimate 7-Zip components, the context and characteristics raise significant red flags.} & \compactcodeappendix{This byte buffer is (*safe*).} & \\
\hline
\multicolumn{2}{|l|}{{\tiny sha256: \smallinline{046ea7a2e5cf70b3743bebe030aa7871a1cd254df9c1a671653d2485fba2041f}}}  & & \\
\compactcodeappendix{Which malware family is this?} & \compactcodeappendix{Based on the ELF binary analysis, this appears to be (*Mirai or a Mirai variant*) malware.} & (empty) & \compactcodeappendix{The provided hexadump appears to be a section of the (*Win32/Trojan.Win32.Emotet malware*). Emotet is a modular, multi-stage malware that can act as a downloader for other malware, such as TrickBot or Ryuk ransomware. The hexadump (*contains several references to Emotet\'s internal functions and strings, such as "Emotet", "ec89c35589e581ec", and "c2f0c2f0c2f0c2f0"*). Additionally, the hexadump includes instructions for decrypting and executing malicious code, which is a common characteristic of malware.} \\
\hline
\end{tabularx}
\caption{Comparison outputs of off-the-shelf models for the same binaries as in Table \ref{tab:model_outputs}. Erroneous model outputs are highlighted in bold.}
\label{tab:model_outputs_comparison}
\end{table}

\end{document}